\documentclass[a4paper,11pt]{article}
\usepackage{amsmath}
\usepackage{amsfonts}
\usepackage{latexsym}
\usepackage{graphicx}
\usepackage{multicol}
\usepackage{times}
\usepackage{latexsym}
\usepackage{enumerate}
\usepackage{epsfig}
\usepackage{amssymb}
\usepackage{amsfonts}
\usepackage[english]{babel}
\usepackage{graphics}
\usepackage{amsbsy}
\usepackage{setspace}
\usepackage{float,placeins}
\usepackage{authblk}
\usepackage{multirow}
\usepackage[hidelinks]{hyperref}
\usepackage{enumitem}
\usepackage{booktabs}
\usepackage{caption}
\usepackage[letterpaper, margin=1in]{geometry}
 
\usepackage[symbol*]{footmisc}

\DefineFNsymbolsTM{otherfnsymbols}{%
  \textasteriskcentered *
  \textsection   \mathsection
  \textbardbl    \|%
}%

\setfnsymbol{otherfnsymbols}

\begin{document}
\author[1]{Roberto Di Mari\thanks{roberto.di.mari@uniroma2.it}}
\author[1]{Roberto Rocci\thanks{roberto.rocci@uniroma2.it}}
\author[2]{Stefano Antonio Gattone\thanks{gattone@unich.it}}
\title{Estimation of clusterwise linear regression models with a shrinkage-like approach}
\affil[1]{DEF, University of Rome Tor Vergata, Italy}
\affil[2]{DiSFPEQ, University G. d'Annunzio, Chieti-Pescara, Italy}

\maketitle

\begin{abstract}
Constrained approaches to maximum likelihood estimation in the context of finite mixtures of normals have been presented in the literature. A fully data-dependent constrained method for maximum likelihood estimation of clusterwise linear regression is proposed, which extends previous work in equivariant data-driven estimation of finite mixtures of Gaussians for classification. The method imposes plausible bounds on the component variances, based on a target value estimated from the data, which we take to be the homoscedastic variance. Nevertheless, the present work does not only focus on classification recovery, but also on how well model parameters are estimated. In particular, the paper sheds light on the shrinkage-like interpretation of the procedure, where the target is the homoscedastic model: this is not only related to how close to the target the estimated scales are, but extends to the estimated clusterwise linear regressions and classification. We show, based on simulation and real-data based results, that our approach yields a final model being the most appropriate-to-the-data compromise between the heteroscedastic model and the homoscedastic model. 
\\

\end{abstract}

\section{Introduction}\label{intro}
Let $\text{y}_{i}$ be the random variable of interest, and let $\mathbf{x}_{i}$ be a vector of $J$ explanatory variables. Whenever a single set of regression coefficients is not adequate for all realizations of the pair $(\text{y}_{i},\mathbf{x}_{i}),$ finite mixture of linear regression models can be used to estimate clusterwise linear regression parameters. 

Let $\text{y}_{i}$ be distributed as a clusterwise linear regression model, that is

\begin{equation}\label{eq:pop}
f(\text{y}_{i}) = {\sum_{g=1}^{G}p_{g}f_{g}(\text{y}_{i}|\mathbf{x}_{i};\sigma_{g}^{2},\pmb{\beta}_{g}}) = {\sum_{g=1}^{G}}p_{g}{\frac{1}{\sqrt{2\pi{\sigma_{g}}^2}}}\exp\bigg(-\frac{(\text{y}_{i}-\mathbf{x}_{i}^{\prime}\pmb{\beta}_{g})^2}{2\sigma_{g}^2}\bigg),
\end{equation}
where $G$ is the total number of clusters and $p_{g},$ $\pmb{\beta}_{g},$ and $\sigma^2_{g}$ are respectively the mixing proportion, the vector of $J$ regression coefficients, and the variance term for the $g$-th cluster. The model in Equation \eqref{eq:pop} is also known under the name of finite mixture of linear regression models, or switching regression model (Quandt, 1972; Quandt and Ramsey, 1978; Kiefer, 1978). 

In addition let us denote the set of parameters to be estimated $\pmb{\psi} \in \pmb{\Psi},$ where \\$\pmb{\psi}=\{(p_{1},\dots,p_{G};\pmb{\beta}_{1},\dots,
\pmb{\beta}_{G};\sigma^{2}_{1},\dots,\sigma^{2}_{G}) \in \mathbb{R}^{G(J + 2) } : p_{1}+\dots+p_{G}=1, p_g \geq 0, \sigma_{g}^{2}>0,$ $g=1, \dots, G\}.$ Unlike finite mixtures of other densities, the parameters of finite mixtures of linear regression models are identified if some mild regularity conditions are met (Hennig, 2000).

The clusterwise linear regression model of Equation \eqref{eq:pop} can naturally serve as a classification model. Based on the model, one computes the posterior membership probabilities as follows
\begin{equation}
\text{p}(g|\text{y}_{i})=\frac{p_{g}f_{g}(\text{y}_{i}|\mathbf{x}_{i};\sigma_{g}^{2},\pmb{\beta}_{g})}
{\sum_{h=1}^{G}p_{h}f_{h}(\text{y}_{i}|\mathbf{x}_{i};\sigma_{h}^{2},\pmb{\beta}_{h})},
\end{equation}
and then classify each observation according, for instance, to fuzzy or crisp classification rules.

Let  $\{(\text{y}_i,\mathbf{x}_{i})\}_n = \{(\text{y}_1,\mathbf{x}_{1}),\dots,(\text{y}_n,\mathbf{x}_{n}) \}$ be a sample of independent observations. In order to estimate $\pmb{\psi},$ one has to maximize the following sample likelihood function

\begin{equation}
\text{L}(\pmb{\psi}) = \prod_{i=1}^{n}\bigg[{\sum_{g=1}^{G}}p_{g}{\frac{1}{\sqrt{2\pi{\sigma_{g}}^2}}}\exp\bigg(-\frac{(\text{y}_{i}-
\pmb{\text{x}}_{i}^{\prime}\pmb{\beta}_{g})^2}{2\sigma_{g}^2}\bigg)\bigg],
\end{equation}

which can be done using iterative procedures like the EM algorithm (Dempster, Laird, and Rubin, 1977). Unfortunately, maximum likelihood (ML) estimation of univariate mixtures of normals - or conditional normals - suffers from the well-known issue of unboundedness of the likelihood function: whenever a sample point coincides with the group's centroid and the relative variance approaches zero, the likelihood function increases without bound (Kiefer and Wolfowitz, 1956; Day, 1969). Hence a global maximum cannot be found.

Kiefer (1978) showed that there is a sequence of consistent, asymptotically efficient and normally distributed estimators for switching regressions with different group-specific variances (heteroscedastic switching regressions). These roots correspond, with probability approaching one, to local maxima in the interior of the parameter space. Yet, although there is a local maximum which is also a consistent root, there is no tool for choosing it among the local maxima. Day (1969) showed that potentially each sample point - or any pair of sample points being sufficiently close together - can generate a singularity in the likelihood function of a mixture with heteroscedastic components. This gives rise to a number of spurious maximizers (McLachlan and Peel, 2000), which, in
a multivariate context, arise from data points being almost coplanar (Ritter, 2014).

The issue of degeneracy can be dealt with by imposing constraints on the component variances. This approach is based on the seminal work of Hathaway (1985), who showed that imposing a lower bound, say $c$, to the ratios of the scale parameters prevents the likelihood function from degeneracy. Although the resulting ML estimator is consistent and the method is equivariant under linear affine transformations of the data - that is, if the data are linearly transformed, the estimated posterior probabilities do not change and the clustering remains unaltered - the proposed constraints are very difficult to apply within iterative procedures like the EM algorithm. Furthermore, the issue of how to choose $c,$ which controls the strength of the constraints, remains open.

Ingrassia (2004) showed that it is sufficient, for Hathaway's constraints to hold, to impose bounds on the component variances. This provides a constrained solution that 1) can be implemented at each iteration of the EM algorithm, and 2) still preserves the monotonicity of the resulting EM (Ingrassia and Rocci, 2007).

We propose a constrained solution to the problem of degeneracy, which extends the one proposed in a recent paper by Rocci, Gattone, and Di Mari (RGD, 2016) for multivariate mixtures of Gaussians, in finite mixtures of linear regression models. This constrained estimation method is characterized by 1) fully data-dependent constraints, 2) affine equivariance of the clustering algorithm under change of scale in the data, and 3) ease of implementation within standard routines (Ingrassia, 2004; Ingrassia and Rocci, 2007). While inheriting such properties from RGD (2016), where the focus was mainly on classification, the extension we consider in this work intrinsically pays attention to how well regression parameters are estimated with the new constrained method, together with looking at how accurately the method classifies observations within clusters. The affine equivariance property, in our context, translates to having a clustering model which yields the same clustering of the data if the variable of interest $\text{y}_i$ is re-scaled. This covers, for instance, cases in which a different scale of the variable of interest is more appropriate for interpreting the regression coefficient than the one the variable was originally measured at. Whereas the simulation study will aim at showing the soundness of the method, the three empirical applications will enlighten the \emph{shrinkage}-like nature of the approach in terms of estimated model parameters and clustering.

The remainder of the paper is organized as follows. In Section \ref{constrained} we briefly review Hathaway's constraints and the sufficient condition in Ingrassia (2004). Section \ref{constreq} is devoted to a description of the proposed methodology and of the estimation algorithm, which is evaluated through a simulation study, presented in Section \ref{simulation}, and three real-data examples, in Section \ref{data}. Section \ref{conclusion} concludes with some final discussion and some ideas for future research.      

\section{Constrained approaches for ML estimation}\label{constrained}
Hathaway (1985) proposed relative constraints on the variances of the kind
\begin{equation}\label{eq:hath}
 \min_{i \neq j} \frac{\sigma_{i}^{2}}{\sigma_{j}^{2}} \geq c \quad \text{with} \quad c \in (0,1]. 
\end{equation}
Hathaway's formulation of the maximum likelihood problem presents a strongly consistent global solution, no singularities, and a smaller number of spurious maxima. Consistency and robustness of estimators of this sort was already pointed out by Huber (1967, 1981).

Ingrassia (2004) formulated a sufficient condition such that Hathaway's constraints hold, which is implementable directly within the EM algorithm, where the covariance matrices are iteratively updated. In a univariate setup, he shows that constraints in ~\eqref{eq:hath}
are satisfied if
\begin{equation}\label{eq:ing1}
 a \leq \sigma^{2}_{g} \leq b, \quad \text{with} \quad g=1, \dots, G,
\end{equation}
where $a$ and $b$ are positive numbers such that $a/b \geq c.$ Complementing the work of Ingrassia (2004), Ingrassia and Rocci (2007) reformulated Ingrassia (2004)'s constraints in such a way that they can be implemented directly at each iteration of the EM algorithm. The conditions under which the proposed constrained algorithm preserves the monotonicity of the unconstrained EM were also stated: the proposed algorithm yields a non-decreasing sequence of the likelihood values, provided that the initial guess $\sigma^{2 (0)}_{g}$ satisfies the constraints.

\section{The proposed methodology}\label{constreq}
\subsection{Affine equivariant constraints}
Starting form the set of constraints of equation ~\eqref{eq:ing1}, let $\bar{\sigma}^{2}$ be a \emph{target} variance. The set of constraints for a clusterwise linear regression context are formulated as follows

$$\sqrt{c} \leq \frac{\sigma^{2}_{g}}{\bar{\sigma}^{2}} \leq \frac{1}{\sqrt{c}},$$
or equivalently
\begin{equation}\label{eq:constr2}
\bar{\sigma}^{2}\sqrt{c} \leq \sigma^{2}_{g} \leq \bar{\sigma}^{2}\frac{1}{\sqrt{c}}.
\end{equation}
It is easy to show that ~\eqref{eq:constr2} implies ~\eqref{eq:hath} - since ~\eqref{eq:constr2} is more stringent than ~\eqref{eq:hath} - while the converse is not necessarily true, that is
$$\frac{\sigma_{g}^{2}}{\sigma_{j}^{2}} = \frac{\sigma_{g}^{2}/\bar{\sigma}^{2}}{\sigma_{j}^{2}/\bar{\sigma}^{2}} \geq \frac{\sqrt{c}}{1/\sqrt{c}} = c.$$

The constraints \eqref{eq:constr2} are affine equivariant (RGD, 2016), and have the effect of shrinking the component variances to $\bar{\sigma}^{2},$ the \emph{target} variance term, and the level of shrinkage is given by the value of $c$: such a formulation makes it possible to reduce the number of tuning constants from two - $(a,b)$ as in Ingrassia (2004)'s proposal - to one - $c.$ Note that for $c=1$, $\hat{\sigma}^{2}_g = \bar{\sigma}^{2},$ whereas for $c\rightarrow 0,$ $\hat{\sigma}^{2}_g$ equals the unconstrained ML estimate. Intuitively, the constraints \eqref{eq:constr2} provide with a way to obtain a model in between a too restrictive model, the homoscedastic, and an ill-conditioned model, the heteroscedastic. In other terms, high scale balance is generally an asset - as it means that there is some unknown transformation of the sample space that transfers the component not too far from the common variance setting - but it has to be traded with fit (Ritter, 2014).

\subsection{Adaptive choice of $c$ and parameter updates}
Selecting $c$ jointly with the mixture parameters by maximizing the likelihood on the entire sample would trivially yield an overfitted scale balance approaching zero (RGD, 2016). Following RGD (2016), we propose a cross-validation strategy in order to let the data decide the optimal scale balance, which can be described according to the following steps.
\begin{enumerate}[leftmargin=*]
\item Select a plausible value for $c \in (0,1].$
\item Obtain a temporary estimate $\hat{\pmb{\psi}}$ for the model parameters using the entire sample, which is used as starting value for the cross-validation procedure.
\item Partition the full data set into a training set, of size $n_{S},$ and a test set, of size $n_{\bar{S}}.$ 
\item Estimate the parameters on the training set using the starting values obtained in step 2. Compute the contribution to the log-likelihood of the test set. 
\item Repeat $K$ times steps 3-4 and sum the contributions of the test sets to the log-likelihood which yields to the so-called cross-validated log-likelihood.
\item Select $c$ which maximizes the cross-validated log-likelihood.
\end{enumerate}

Smyth (1996; 2000) advocates the use of the test set log-likelihood for selecting the number of mixture components. The rationale is that it can be shown to be an unbiased estimator (within a constant) of the Kullback-Leibler divergence between the \emph{truth} and the model under consideration. As large test sets are hardly available, the cross-validate log-likelihood can be used to estimate the test set log-likelihood. In our case - like in Smyth's case (1996) - given the model parameters, the cross-validated log-likelihood is a function of $c$ only, and maximizing it with respect to $c,$ given the other model parameters, would handle the issue of overfitting as training and test sets are independent (Arlot and Celisse, 2000).

The updates for the quantities of interest are obtained as follows. For a chosen value of the constant $c,$ the expectation step (E-step) at the $(k+1)$-th iteration produces a guess for the quantity
\begin{equation}
\text{z}_{ig}^{(k+1)}(\text{y}_{i};\pmb{\psi}^{(k)})=\frac{p_{g}f_{g}(\text{y}_{i};\pmb{\beta}_{g}^{(k)},\sigma^{2(k)}_{g})}
{\sum_{h=1}^{G}p_{h}f_{h}(\text{y}_{i};\pmb{\beta}_{g}^{(k)},\sigma^{2(k)}_{g})},
\end{equation}
where $i=1,\dots,n$ and $g=1,\dots,G.$ Using the computed quantities from step E, the maximization step (M-step) involves the following closed-form updates:
\begin{equation}
p_{g}^{(k+1)}=\frac{1}{n}\sum_{i=1}^{n}\text{z}_{ig}^{(k+1)},
\end{equation}
\begin{equation}
\pmb{\beta}_{g}^{(k+1)}=(\sum_{i=1}^{n}\text{z}_{ig}^{(k+1)}\pmb{\text{x}}_{i}\pmb{\text{x}}_{i}^{\prime})^{-1}\sum_{i=1}^{n}
\text{z}_{ig}^{(k+1)}\pmb{\text{x}}_{i}\text{y}_{i},
\end{equation}
\begin{equation}\label{eq:variance}
\sigma^{2(k+1)}_{g}=\frac{\sum_{i=1}^{n}z_{ig}^{(k)}(\text{y}_{i}-\pmb{\text{x}}_{i}^{\prime}\pmb{\beta}_{g})^{2}}{\sum_{i=1}^{n}z_{ig}^{(k)}}.
\end{equation}

At each M-step, the final update for the variance is given by the following expression
\begin{equation}\label{eq:rule}
\sigma_{cg}^{2(k+1)} = \min(\bar{\sigma}^{2}\frac{1}{\sqrt{c}},\max(\bar{\sigma}^{2}\sqrt{c},\sigma^{2(k+1)}_{g})), \quad \text{for $g=1,\dots,G$},
\end{equation}
where $\sigma_{cg}^2(k+1)$ indicates the $g$-th component constrained variance at $k+1$-th iteration.

Constraining the components to common scale yields instead the following update for the variance
\begin{equation}\label{eq:homovar}
\bar{\sigma}^{2(k+1)}=\frac{1}{n}\sum_{i=1}^{n}\sum_{g=1}^{G}z_{ig}^{(k)}(\text{y}_{i}-\pmb{\text{x}}_{i}^{\prime}\pmb{\beta}_{g})^{2},
\end{equation}
which is the natural candidate for \emph{target} for our shrinkage-like procedure.

In this constrained version of the EM, provided the initial guess satisfies the constraints, the algorithm is monotonic and the complete log-likelihood is maximized.

\section{Numerical study}\label{simulation}
\subsection{Design}
A simulation study was conducted in order to evaluate the quality of the parameter estimates of our constrained algorithm. The equivariant data-driven constrained algorithm (ConC) was compared with the unconstrained algorithm with common (homoscedastic) component-scales (HomN), and the unconstrained algorithm with different (heretoscedastic) component-scales (HetN). The target measures used for the comparisons were average Mean Squared Errors (MSE) of the regression coefficients (averaged across regressors and groups), and MSE of the component variances (averaged across groups). These measures will allow us to state the precision of the estimates. We measured how well the algorithms were able to classify sample units within clusters through the adjusted \emph{Rand} index (Adj-Rand, Hubert and Arabie, 1985).

The data were generated from a clusterwise linear regression with 3 regressors and intercept, with 2 and 3 components and sample sizes of 100 and 200. The class proportions considered were, respectively, $(0.5,0.5)^\prime$ and $(0.2,0.8)^\prime,$ and $(0.2,0.4,0.4)^\prime$ and $(0.2,0.3,0.5)^\prime.$ Regressors have been drawn from 3 independent standard normals, whereas regression coefficients have been drawn from U$(-1.5,1.5)$ and intercepts are $(4,9)^{\prime}$ and $(4,9,16)^{\prime}$ for the 2-component and 3-component models respectively. The component variances have been drawn from Inv-Gamma$(3,1)$. 

For each of the 8 combinations $\{n,p_{g}\},$ we generated 250 samples: for each sample and each algorithm - HomN, HetN, ConC (our proposal) - we select the best solution (highest likelihood) out of 10 random starts. The simulated-data analysis was conducted in MATLAB. 

\subsection{Results}

Table \ref{tab:100obs} displays the results for $n=100.$ 2 groups and even class proportions (the $(0.5,0.5)^\prime$ case) is the most favorable condition for the unconstrained algorithms in terms of regression parameters' MSE. As we consider a larger number of classes ($G=3$), ConC yields an MSE for the betas which is 34\% smaller than the ones reported for HetN and HomN. The percentage difference nearly doubles when class proportions are uneven (the $(0.2,0.3,0.5)^\prime$ case). Similar results are reported for the variances' MSE, with exception of the $(0.5,0.5)^\prime$ case. ConC delivers the most accurate classification in all 4 conditions.

Increasing the sample size to 200 (Table \ref{tab:200obs}) improves the performance of all methods, especially HetN and ConC. Similarly to what has been reported for $n=100,$ for the 3-group cases ConC has an MSE for the betas at least 66\% smaller than the unconstrained methods - the difference is even larger for the variances' MSE. The Adj-Rand reported for ConC is the closest to 1, with exception of the $(0.5,0.5)^\prime$ case, where HetN does slightly better than ConC (0.9784, compared to 0.9777 for ConC).

\FloatBarrier

\begin{table}
\centering
 \begin{tabular}{lcccccccccc}
\hline \hline
Algorithm && Avg MSE $\hat{\pmb{\beta}}$  && Avg MSE $\hat{\pmb{\sigma}}^{2}$ && $\text{Adj-Rand}$ && time && $c$ \\
 \hline

&&\multicolumn{9}{c}{Mixing proportions $(0.5,0.5)$}\\
     &&	       &&        &&	   &&        &&        \\     
	
HomN && 0.0159 && 0.0367 && 0.9399 && 0.0485 &&   -    \\
HetN && 0.0135 && 0.0076 && 0.9616 && 0.0492 &&   -    \\
ConC && 0.0135 && 0.0082 && 0.9626 && 0.7694 && 0.4824 \\

\cmidrule{3-11}

&&\multicolumn{9}{c}{Mixing proportions $(0.2,0.8)$}\\
     &&	       &&        &&	   &&        &&        \\

HomN && 0.0402 && 0.0510 && 0.9648 && 0.0567 &&  -     \\
HetN && 0.0522 && 0.0439 && 0.9672 && 0.0525 &&  -     \\
ConC && 0.0416 && 0.0238 && 0.9718 && 0.8026 && 0.5102 \\

\cmidrule{3-11}

&&\multicolumn{9}{c}{Mixing proportions $(0.2,0.4,0.4)$}\\
     &&	       &&        &&	   &&        &&        \\

HomN && 0.3192 && 0.1107 && 0.9403 && 0.1763 &&  -     \\
HetN && 0.3201 && 0.1665 && 0.9318 && 0.1804 &&  -     \\
ConC && 0.2117 && 0.0916 && 0.9481 && 1.4315 && 0.2875 \\

\cmidrule{3-11}

&&\multicolumn{9}{c}{Mixing proportions $(0.2,0.3,0.5)$}\\
     &&	       &&        &&	   &&        &&        \\

HomN && 0.2482 && 0.0899 && 0.9546 && 0.1868 &&   -    \\
HetN && 0.3226 && 0.1226 && 0.9579 && 0.1446 &&   -    \\
ConC && 0.1051 && 0.0462 && 0.9688 && 1.2933 && 0.2436 \\

\hline \hline 
\end{tabular}
\caption{250 samples, $n=100,$ 10 random starts, 3 regressors and intercept. Values averaged across samples. \label{tab:100obs}}
\end{table}

\begin{table}
\centering
\begin{tabular}{lcccccccccc}
\hline \hline
Algorithm && Avg MSE $\hat{\pmb{\beta}}$  && Avg MSE $\hat{\pmb{\sigma}}^{2}$ && $\text{Adj-Rand}$ && time && $c$ \\
 \hline

&&\multicolumn{9}{c}{Mixing proportions $(0.5,0.5)$}\\
     &&	       &&        &&	   &&        &&        \\     
	
HomN && 0.0070 && 0.0284 && 0.9574 && 0.0954 &&    -   \\
HetN && 0.0057 && 0.0032 && 0.9784 && 0.0977 &&    -   \\
ConC && 0.0057 && 0.0033 && 0.9777 && 2.2902 && 0.4884 \\

\cmidrule{3-11}

&&\multicolumn{9}{c}{Mixing proportions $(0.2,0.8)$}\\
     &&	       &&        &&	   &&        &&        \\

HomN && 0.0121 && 0.0384 && 0.9762 && 0.1116 &&   -    \\
HetN && 0.0137 && 0.0083 && 0.9839 && 0.0954 &&   -    \\
ConC && 0.0109 && 0.0059 && 0.9841 && 2.2682 && 0.4633 \\

\cmidrule{3-11}

&&\multicolumn{9}{c}{Mixing proportions $(0.2,0.4,0.4)$}\\
     &&	       &&        &&	   &&        &&        \\

HomN && 0.0553 && 0.0724 && 0.9716 && 0.3251 &&   -    \\
HetN && 0.0399 && 0.0208 && 0.9835 && 0.2951 &&   -    \\
ConC && 0.0103 && 0.0075 && 0.9856 && 3.4715 && 0.2768 \\

\cmidrule{3-11}

&&\multicolumn{9}{c}{Mixing proportions $(0.2,0.3,0.5)$}\\
     &&	       &&        &&	   &&        &&        \\

HomN && 0.3245 && 0.1476 && 0.9637 && 0.4227 &&    -   \\
HetN && 0.0630 && 0.0234 && 0.9838 && 0.2868 &&    -   \\
ConC && 0.0152 && 0.0116 && 0.9862 && 3.6266 && 0.2067 \\

\hline \hline 
\end{tabular}
\caption{250 samples, $n=200,$ 10 random starts, 3 regressors and intercept. Values averaged across samples. \label{tab:200obs}} 
\end{table}

\FloatBarrier



\section{Three real-data applications}\label{data}

The aim is to show, through the three real-data applications we present in this section, that the method works well in terms of quality of model parameter estimates and classification. Whereas one can compare true with estimated parameter values in a simulation study (where the data generating process is known), this is not possible in real-data applications, where possibly also the number of clusters in the sample is unknown. Having this in mind, we estimated a clusterwise linear regression, using the 3 methods under comparison, on the \emph{CEO} data set (\url{http://lib.stat.cmu.edu/DASL/DataArchive.html}), with 2, 3 and 4 components, assessing the plausibility of the estimated regression lines. We carried out a similar exercise on the \emph{Temperature} data set (Long, 1972; available at \url{http://rcarbonneau.com/ClusterwiseRegressionDatasets/data_USTemperatures.txt}),\\ where we analyzed and compared clusterwise linear regression models with 2, 3, 4 and 5 components. 

Finally, we estimated a 3-component clusterwise linear regression model on Fisher's Iris data, using petal width as dependent variable and sepal width as explanatory variable, in order to assess the clusters recovery. 

In all applications the estimated groups are ordered from the smallest to the largest in terms of cluster size. The empirical analysis was conducted in R.

\subsection{\emph{CEO} data}
This data set has a well-known structure, although a \emph{true} number of clusters is not available. It contains 59 records for some U.S. small-companies' CEO salaries (dependent variable), and CEO ages (independent variable). 

\begin{center}
\begin{figure}
	\includegraphics[scale=0.9]{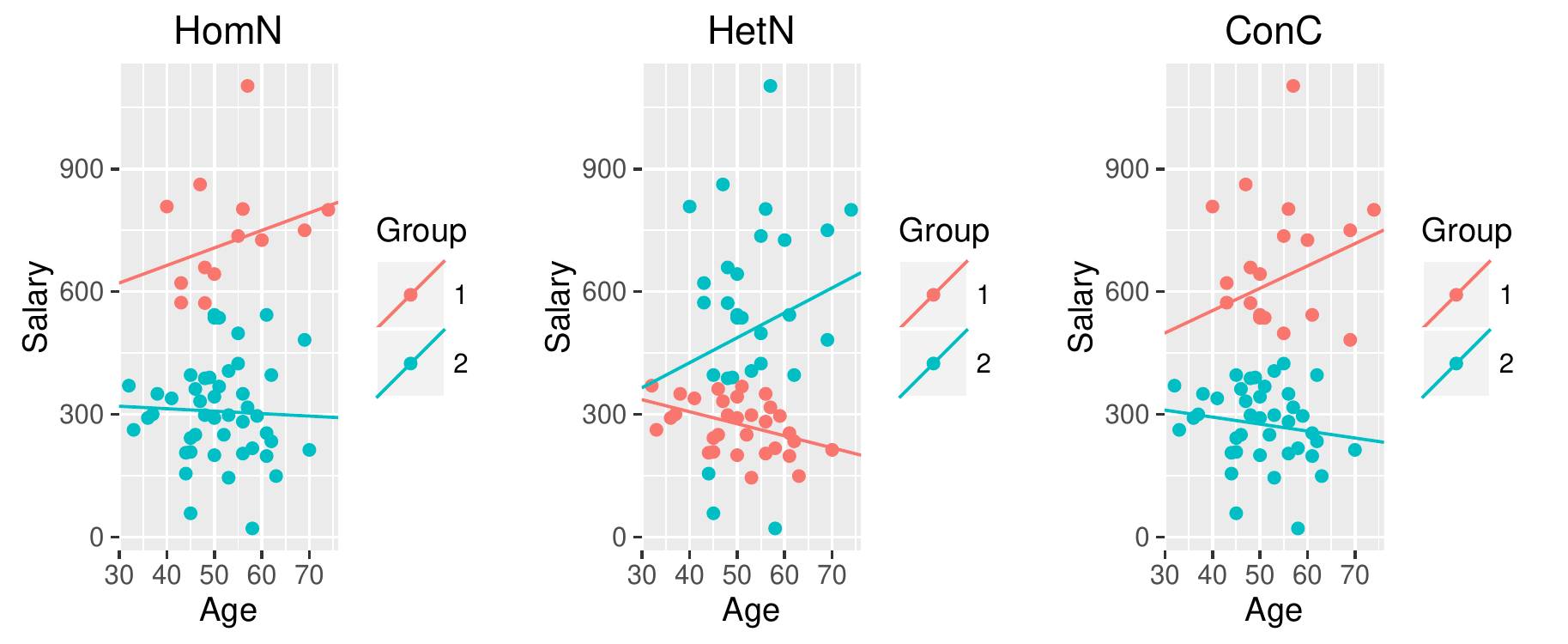}
	\caption{\emph{CEO} data. Best solutions out of 50 random starts, $G= 2$.
	 $K = n/5$, and test set size = $n/10.$}
	\label{fig:ceo2}
\end{figure}
\end{center}

\begin{center}
\begin{figure}
\includegraphics[scale=0.9]{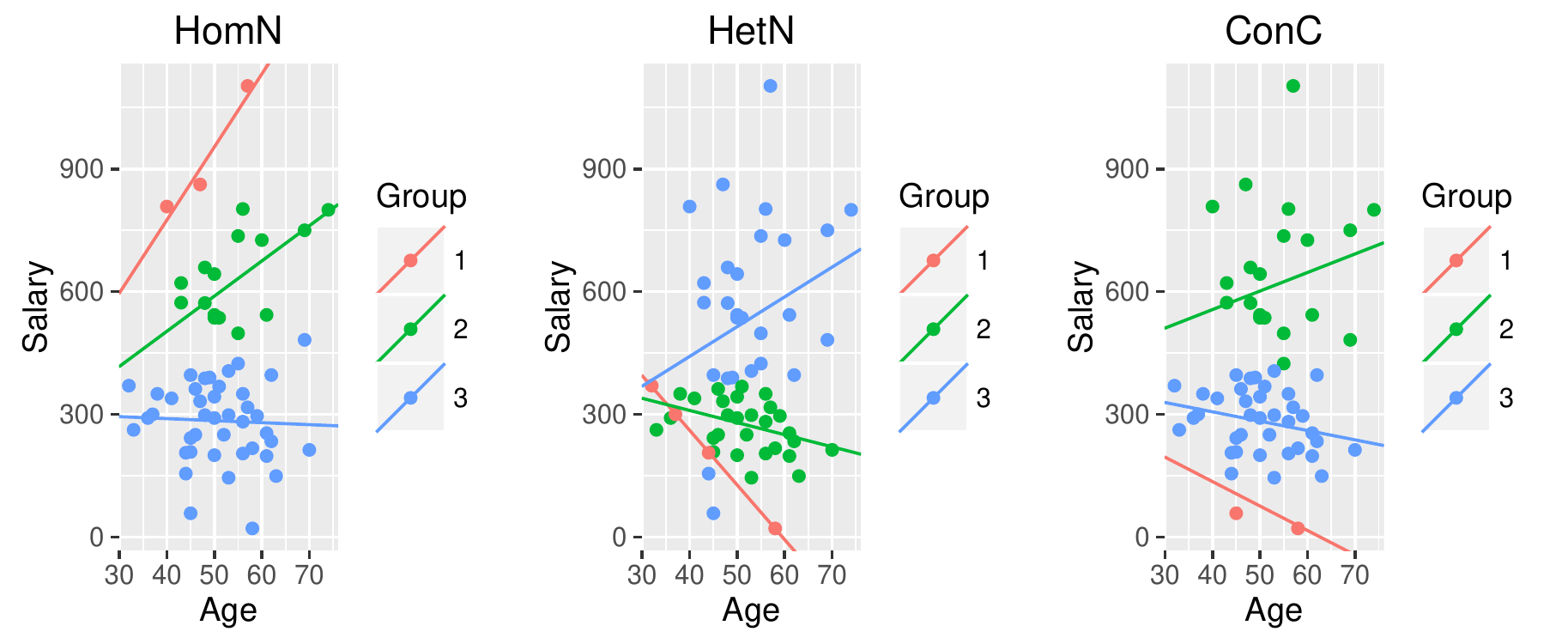}
\caption{\emph{CEO} data. Best solutions out of 50 random starts, $G= 3$. $K = n/5$, and test set size = $n/10.$}
\label{fig:ceo3}
\end{figure}
\end{center}

\begin{center}
\begin{figure}
\includegraphics[scale=0.9]{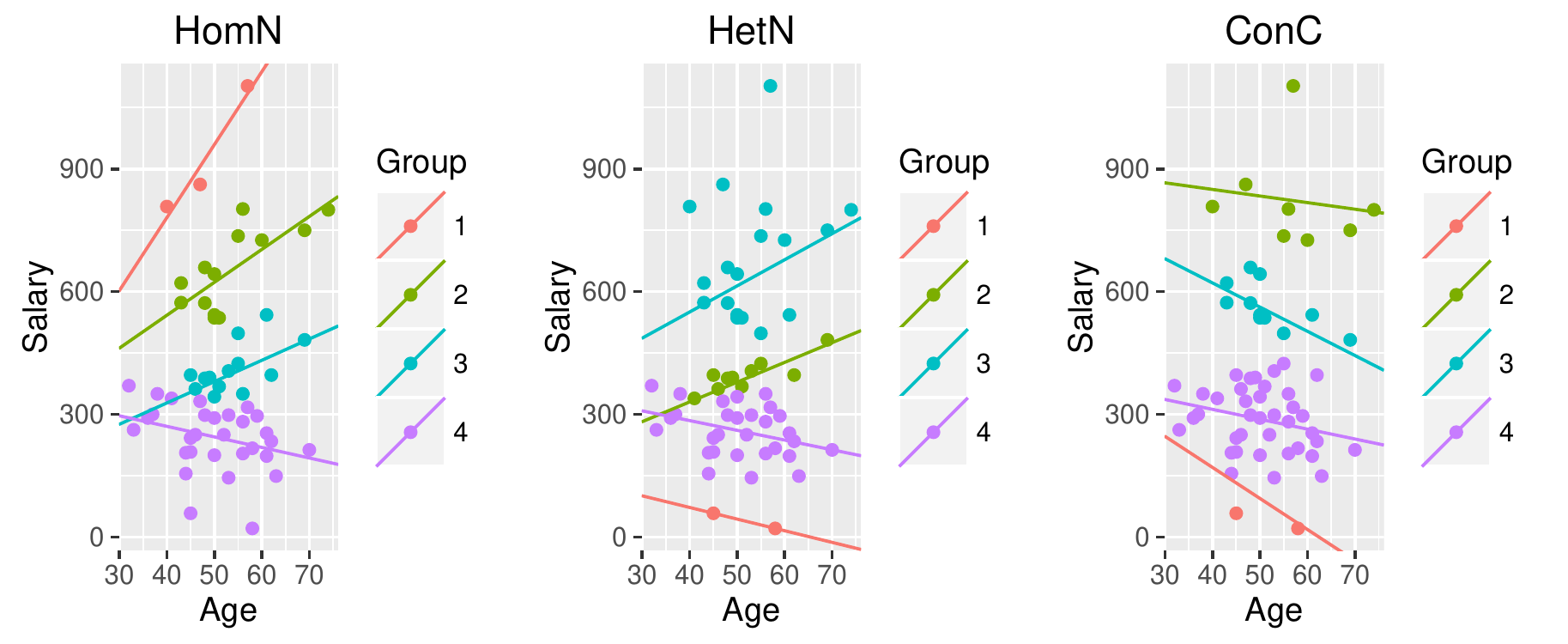}
\captionof{figure}{\emph{CEO} data. Best solutions out of 50 random starts, $G= 4$. $K = n/5$, and test set size = $n/10.$}
\label{fig:ceo4}
\end{figure}
\end{center}

Bagirov, Ugon, and Mirzayeva (2013) fitted a 2-component and a 4-component clusterwise linear regression, whereas Carbonneau, Caporossi, and Hansen (2011) focused on the perhaps most intuitive 2-component setup. We fitted respectively 2-component (Figure \ref{fig:ceo2}), 3-component (Figure \ref{fig:ceo3}), and 4-component (Figure \ref{fig:ceo4}) clusterwise linear regressions, and graphically compared the regression lines and crisp classifications obtained.   
\FloatBarrier

\FloatBarrier	

The 2-class solution (Figure \ref{fig:ceo2}) would be favored by HomN in terms of BIC, whereas HetN would select a 4-class model (Figure \ref{fig:ceo4}) as the one minimizing the BIC.

As shown in Figure \ref{fig:ceo2}, HomN and HetN deliver different solutions in terms of both estimated regression lines and clustering. Interestingly, ConC, with a selected $c$ of about 0.3, yields a solution which is in between homoscedasticity and heteroscedasticity: this is the case not only in terms of estimated component-scales, as expected, but also in terms of fitted regression lines and clustering. In other words, the proposed procedure has the very nice feature of letting the data decide the most appropriate model as the most suited-to-the-data compromise between homoscedasticity and heteroscedasticity. 

A similar shrinkage-like solution, in between homoscedasticity and heteroscedasticity, is delivered also with 3 (Figure \ref{fig:ceo3}) and 4 mixture components (Figure \ref{fig:ceo4}). In particular, the ConC's solution with 3 components is closer to HetN's than HomN's ($c\approx$0.15), although we observe some departure, especially in the first (red) component: indeed HetN red group's regression line crosses the 4 units assigned to that component. This would signal the spurious nature of the solution delivered by HetN. On the other hand, HomN's solution, especially for the first component, seems to be driven by the unit with the highest salary.

As we turn to the 4 components case, ConC yields results closer to HomN ($c\approx$0.9). In the latter case we observe a solution for HetN which is very likely to be spurious, as the first component's regression line (in red) is aligned with two data points, and the relative component variance is relatively very small (the scale ratio between first and second component is $<10^{-11}$). As for the 3-component case, HomN first component's solution seems to be driven by the unit with the highest salary. 

\subsection{\emph{Temperature} data}

This data set concerns average minimum temperatures in 56 US cities in January, including latitude and longitude of each city\footnote{The time span considered for taking the average is January 1930 - January 1960. For further information on how average minimum temperatures are obtained please refer to Peixoto, 1990}. Among the others who already analyzed these data, Peixoto (1990) fitted a polynomial regression of temperature on latitude and a cubic polynomial in longitude to this data set, whereas Carbonneau and co-authors (2011) fitted a 2-component clusterwise linear regression of temperature on latitude and longitude. We fitted a clusterwise linear regression model of temperature on latitude and longitude with respectively 2 (Table \ref{tab:restemp2}), 3 (Table \ref{tab:restemp3}), 4 (Table \ref{tab:restemp4}), and 5 components (Table \ref{tab:restemp5}). For obvious reasons, latitude is more informative than longitude to determine the final clustering. This is why we will plot the clustering results focusing on temperatures and latitudes only.    
\FloatBarrier
\begin{table}[h!]
\centering
\begin{tabular}{lccc}
\hline \hline
&&\multicolumn{2}{c}{HomN}\\
\cmidrule{3-4}
$p_g$&&0.2371 	 & 	0.7629 	\\ 
intercept &&66.4905 	 & 	142.2348 	\\ 
$\beta_{1g}$ &&-1.7204 	 & 	-2.5182 	\\ 
$\beta_{2g}$ &&0.3353 	 & 	-0.2249 	\\ 
$\sigma^{2}_{g}$ && 3.5602 	 & 	3.5602 	 \\
$c$ &&- 	 & 	- 	 \\
   &&  	 &	 	 \\\hline
&&\multicolumn{2}{c}{HetN} \\
\cmidrule{3-4}
$p_g$&& 0.2626 	 & 	0.7374 	 \\ 
intercept && 74.5325 	 & 	150.6175 	\\ 
$\beta_{1g}$ && -1.9829 	 & 	-2.5806 	\\ 
$\beta_{2g}$ && 0.3434 	 & 	-0.2945 	 \\ 
$\sigma^{2}_{g}$ && 6.0948 	 & 	2.7499 	 \\ 
$c$&&- 	 & 	- 	 \\ 
   &&  	 &	  	 	 \\ \hline
&&\multicolumn{2}{c}{ConC} \\
\cmidrule{3-4}
$p_g$&& 0.2647 	 & 	0.7353 	\\ 
intercept && 74.8657 	 & 	150.5268 	\\ 
$\beta_{1g}$ && -1.9945 	 & 	-2.5791 	\\ 
$\beta_{2g}$ && 0.3460 	 & 	-0.2939 	\\ 
$\sigma^{2}_{g}$ && 5.6950 	 & 	2.7568 	\\  
$c$ && 0.1527 & 0.1527 \\
\hline \hline 

\end{tabular}
\caption{\emph{Temperature} data. Best solutions out of 100 random starts, $G= 2$. $K = n/5$, and test set size = $n/10.$ \label{tab:restemp2}}
\end{table}

\begin{table}[h!]
\centering
\begin{tabular}{lcccc}
\hline \hline
&&\multicolumn{3}{c}{HomN}\\
\cmidrule{3-5}
$p_g$&& 0.1091 	 & 	0.1933 	 & 	0.6975 \\ 
intercept && 52.0354 	 & 	96.7770 	 & 	150.5393 \\ 
$\beta_{1g}$&&-1.2531 	 &  -2.4057 	 & 	-2.5779 	\\ 
$\beta_{2g}$&&0.319 	 & 	0.2735 	 & 	-0.2940 	 \\ 
$\sigma^{2}_{g}$&& 2.8466 	 & 	2.8466 	 & 	2.8466 	\\ 
$c$ &&- 	 & 	- 	 & 	- 	  \\
   &&  	 &	 	 &  	 \\\hline
&&\multicolumn{3}{c}{HetN} \\
\cmidrule{3-5}
$p_g$&& 0.1841 	 & 	0.2727 	 & 	0.5432 	 \\ 
intercept && 148.7525 	 & 	75.1858 	 & 	155.3267 	 \\ 
$\beta_{1g}$&& -2.9327 	 & 	-1.9155 	 & 	-2.3968 	\\ 
$\beta_{2g}$&& -0.1350 	 & 	0.3031 	 & 	-0.4282 	\\ 
$\sigma^{2}_{g}$&& 0.4023 	 & 	6.7563 	 & 	1.7041 	 \\ 

$c$&&- 	 & 	- 	 & 	- 	 \\ 
   &&  	 &	  	 	 &  	 \\ \hline
&&\multicolumn{3}{c}{ConC} \\
\cmidrule{3-5} 
$p_g$&&0.1806 	 & 	0.2635 	 & 	0.5559 	 \\ 
intercept &&118.0823 	 & 	63.4605 	 & 	159.4095 	\\ 
$\beta_{1g}$&&-2.0272 	 & 	-1.8391 	 & 	-2.5445 	\\ 
$\beta_{2g}$&&-0.1385 	 & 	0.4038 	 & 	-0.4119 	 \\ 
$\sigma^{2}_{g}$&&2.1395 	 & 	3.7875 	 & 	2.1395 	\\ 
$c$&& 0.3191  &  0.3191  & 0.3191  \\
\hline \hline 

\end{tabular}
\caption{\emph{Temperature} data. Best solutions out of 100 random starts, $G= 3$. $K = n/5$, and test set size = $n/10.$ \label{tab:restemp3}}
\end{table}

\begin{table}[h!]
\centering
\begin{tabular}{lccccc}
\hline \hline
&&\multicolumn{4}{c}{HomN}\\
\cmidrule{3-6}
$p_g$&& 0.0856 	 & 	0.1881 	 & 	0.2057 	 & 	0.5206 	 \\ 
intercept && 51.8842 	 & 	117.1396 	 & 	83.4083 	 & 	159.6096 \\ 
$\beta_{1g}$ &&-1.1266 	 & 	-1.9904 	 & 	-2.1729  & 	-2.5503 \\ 
$\beta_{2g}$ && 0.2758 	 & 	-0.1430 	 & 	0.3186 	 & 	-0.4123 \\ 
$\sigma^{2}_g$&&1.8003 	 & 	1.8003 	 & 	1.8003 	 & 	1.8003 	 \\ 
$c$ &&- 	 & 	- 	 & 	- 	 & 	- 	 \\
   &&  	 &	 	 &  	 & 	 	 \\\hline
&&\multicolumn{4}{c}{HetN} \\
\cmidrule{3-6}
$p_g$&& 0.1459 	 & 	0.2074 	 & 	0.2148 	 & 	0.4319 	 \\ 
intercept && 112.1847 	 & 	62.8423 	 & 	143.6033 	 & 	162.6549 	 \\ 
$\beta_{1g}$ && -1.7960 	 & 	-1.6773 	 & 	-2.8471 	 & 	-2.5137 	 \\ 
$\beta_{2g}$ && -0.1775 	 & 	0.3498 	 & 	-0.1177 	 & 	-0.4610 	 \\ 
$\sigma^{2}_g$ && 0.3626 	 & 	4.4269 	& 	0.3000 	 & 	1.8447 	 \\ 
$c$&&- 	 & 	- 	 & 	- 	 & 	- 	 \\ 
   &&  	 &	  	 	 &  	 & 	 	 \\ \hline
&&\multicolumn{4}{c}{ConC} \\
\cmidrule{3-6} 
$p_g$&& 0.1922 	 & 	0.2387 	 & 	0.2562 	 & 	0.3129 	 \\ 
intercept && 65.9919 	 & 	179.3683 	 &	114.8658 	 & 	140.3715 	\\ 
$\beta_{2g}$ && -1.6827 	 & 	-2.7064 	 & 	-1.9692 	 & 	-2.7506 	\\ 
$\beta_{1g}$ && 0.3263 	 & 	-0.5678 	 & 	-0.1328 	 & 	-0.1229 	 \\ 
$\sigma^{2}_g $ && 3.5915 	 & 	1.1907 	 & 	1.1907 	 & 	1.1907 	 \\ 
$c$&&0.1099 	& 	0.1099 	& 0.1099	 & 	0.1099	\\ 
\hline \hline 

\end{tabular}
\caption{\emph{Temperature} data. Best solutions out of 100 random starts, $G= 4$. $K = n/5$, and test set size = $n/10.$ \label{tab:restemp4}}
\end{table}

\begin{table}[h!]
\centering
\begin{tabular}{lcccccc}
\hline \hline
&&\multicolumn{5}{c}{HomN}\\
\cmidrule{3-7} 
$p_{g}$&&0.0881 	 & 	0.1275 	 & 	0.2117 	 & 	0.2444 	 & 	0.3284 	 \\ 
intercept && 53.2996 	 & 	53.1467 	 & 	179.4572 	 & 	114.5672 	 & 	140.4554 	 \\ 
$\beta_{1g}$ && -0.8048 	 & 	-1.1204 	 & 	-2.7236 	 & 	-1.9806 	 & 	-2.7523 	 \\ 
$\beta_{2g}$ && 0.0347 	 & 	0.2640 	 & 	-0.5639 	 & 	-0.1257 	 & 	-0.1224 	 \\ 
$\sigma^{2}_{g}$ && 0.9200 	 & 	0.9200 	 & 	0.9200 	 & 	0.9200 	 & 	0.9200 	 \\ 
$c$ && - 	 &	- 	 & 	- 	 & 	- 	 & 	- 	 \\ 
   &&  	 &	 	 & 	 	 &  	 & 	 	 \\\hline
&&\multicolumn{5}{c}{HetN} \\
\cmidrule{3-7}
$p_{g}$&& 0.0536 	 & 	0.1180 	 & 	0.1515 	 &	0.2168 	 & 	0.4600 	 \\ 
intercept && 130.2868 	 & 	52.4339 	 & 	109.4192 	 & 	143.6856 	 & 	155.3994 	 \\ 
$\beta_{1g}$ && -2.5978 	 & 	-1.1145 	 & 	-2.4808 	 & 	-2.8487 	 &	-2.3851 	 \\ 
$\beta_{2g}$ && -0.0491 	 & 	0.2673 	 & 	0.1908 	 & 	-0.1179 	 & 	-0.4358 	\\ 
$\sigma^{2}_{g}$ && $10^{-10}$ 	 & 	1.1011 	 & 	3.8704 	& 	0.2956 	 & 	1.7691 	 \\ 
$c$ && - 	 & 	- 	 & 	- 	 & 	- 	 & 	- 	 \\ 
   &&  	 &	 	 & 	 	 &  	 & 	 	 \\ \hline
&&\multicolumn{5}{c}{ConC} \\
\cmidrule{3-7}
$p_g$ && 0.0746 	 & 	0.1309 	 & 	0.2403 	 & 	0.2639 	 & 	0.2903 	 \\ 
intercept && 26.6487 	 & 	52.9665 	 & 	174.4789 	 & 	111.1235 	 & 	139.2966 	\\ 
$\beta_{1g}$ &&-0.6535 	 & 	-1.1149 	 & 	-2.7457 	 & 	-1.7887 	 & 	-2.7735 	\\ 
$\beta_{2g}$ && 0.2028 	 & 	0.2635 	 & 	-0.5017 	 & 	-0.1729 	 & 	-0.1010 	 \\ 
$\sigma^{2}_{g}$ && 0.8116 	 & 	1.0558 	 & 	1.1956 	 & 	0.8939 	 & 	0.8116 	 \\ 
$c$&& 0.4608 	 & 	0.4608 	 & 	0.4608 	 & 	0.4608 	 & 	0.4608\\

\hline \hline 

\end{tabular}
\caption{\emph{Temperature} data. Best solutions out of 100 random starts, $G= 5$. $K = n/5$, and test set size = $n/10.$ \label{tab:restemp5}}
\end{table}

\FloatBarrier

\begin{center}
\begin{figure}[h!]	
	\includegraphics[scale=0.9]{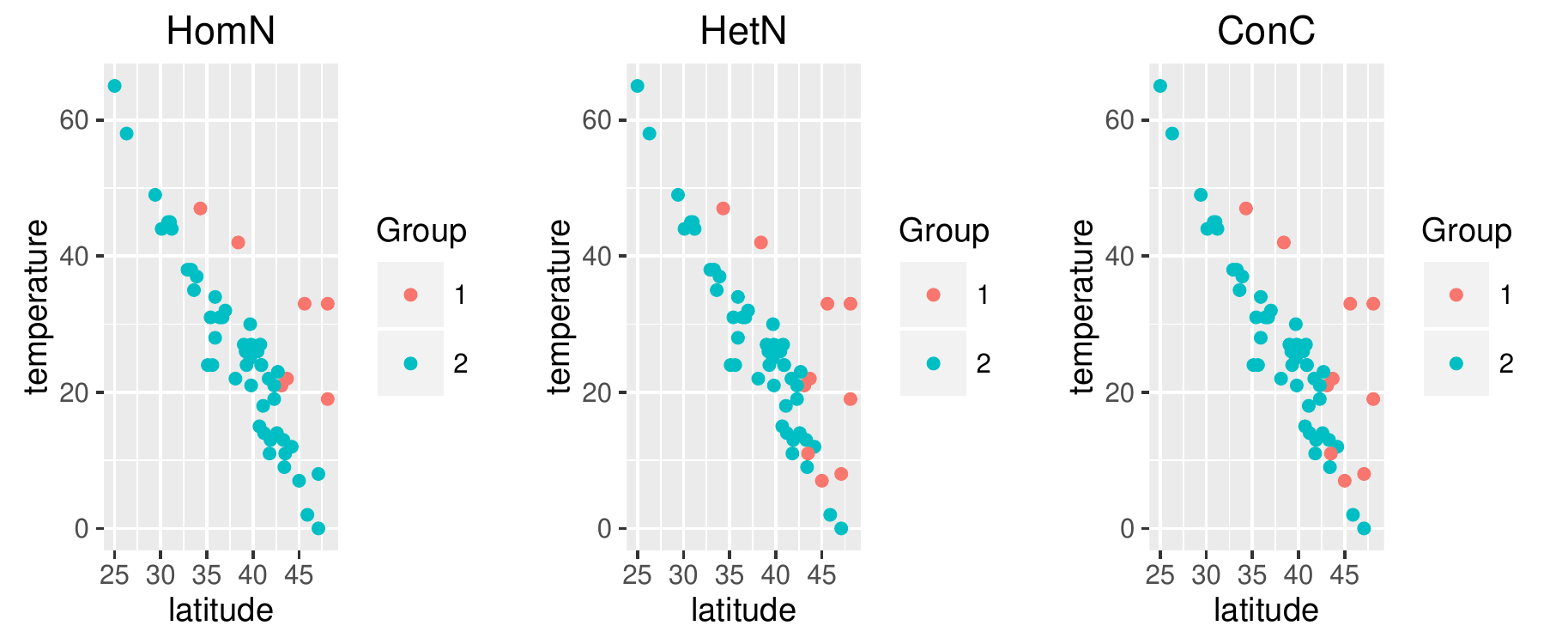}
	\caption{\emph{Temperature} data. Best solutions out of 100 random starts, $G= 2$. $K = n/5$, and test set size = $n/10.$}
	\label{fig:temp2}
\end{figure}
\end{center}


\begin{center}
\begin{figure}[h!]
\includegraphics[scale=0.9]{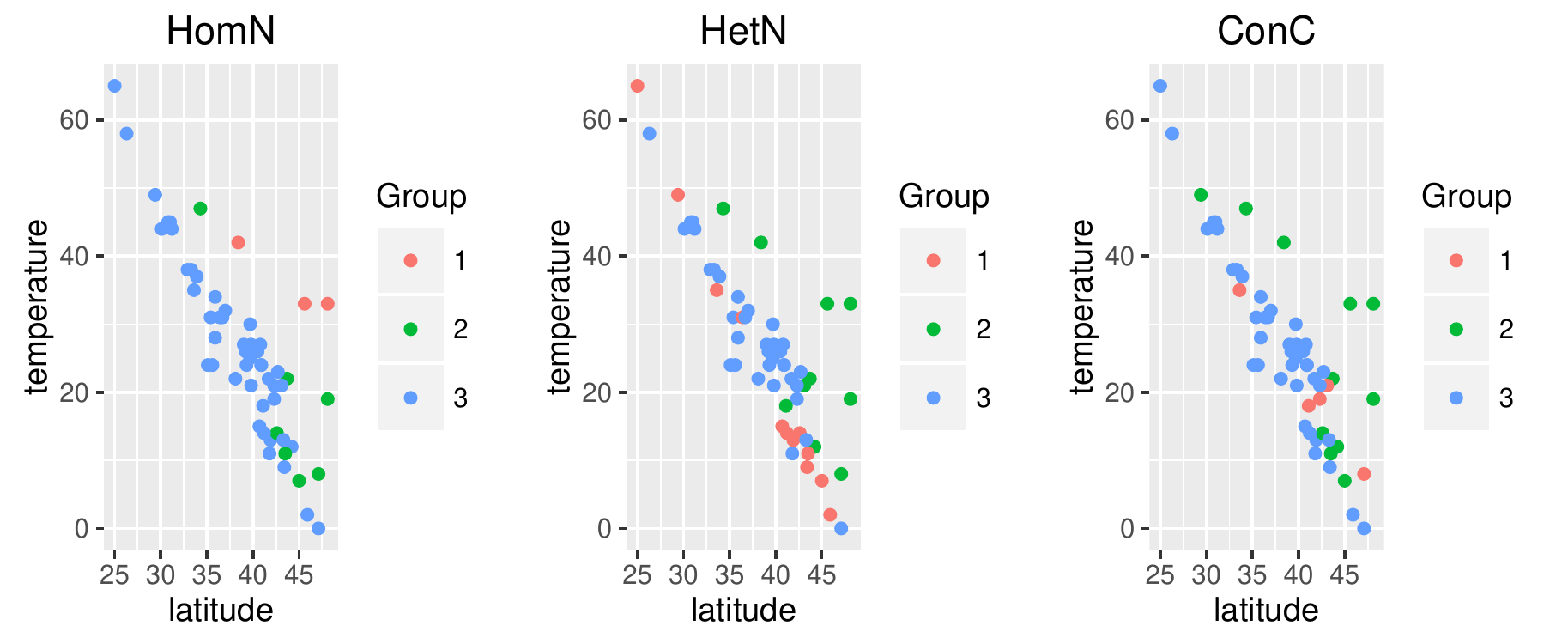}
\caption{\emph{Temperature} data. Best solutions out of 100 random starts, $G= 3$. $K = n/5$, and test set size = $n/10.$}
\label{fig:temp3}
\end{figure}
\end{center}


\begin{center}
\begin{figure}[h!]
\includegraphics[scale=0.9]{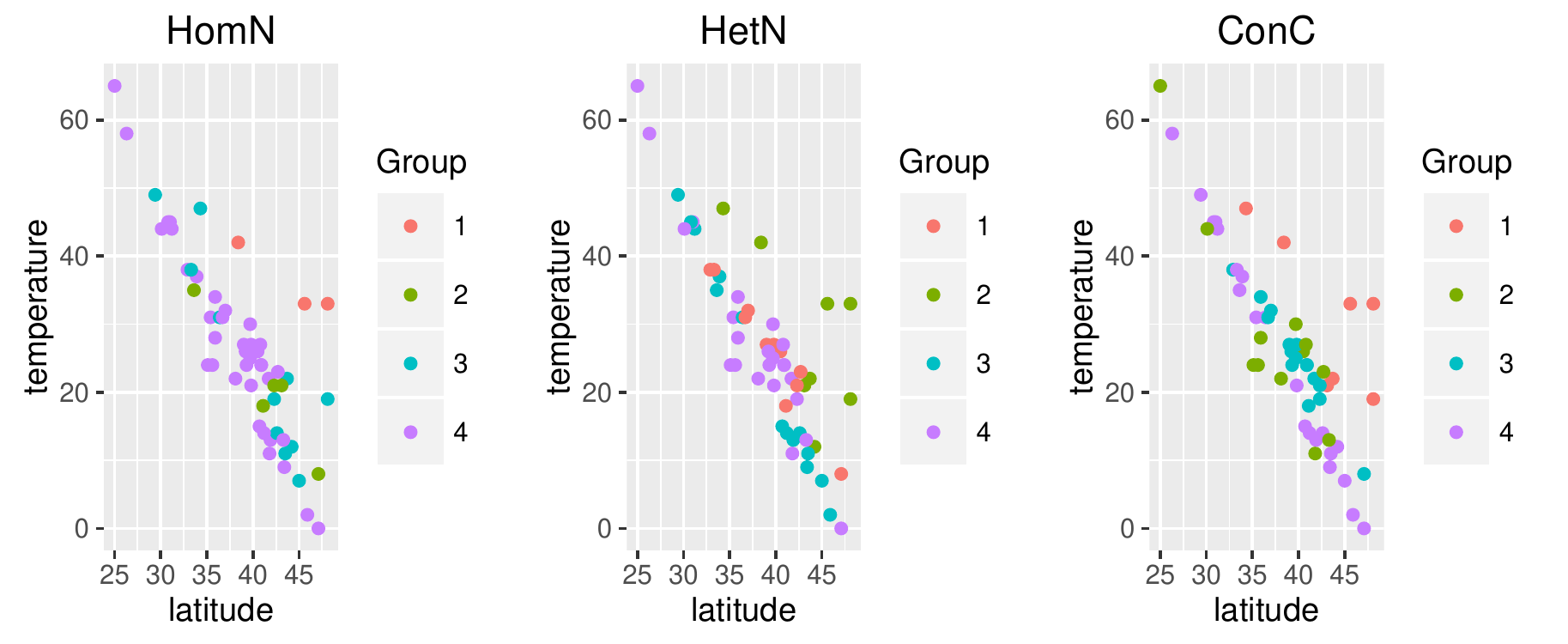}
\caption{\emph{Temperature} data. Best solutions out of 100 random starts, $G= 4$. $K = n/5$, and test set size = $n/10.$}
\label{fig:temp4}
\end{figure}
\end{center}


\begin{center}
\begin{figure}[h!]
\includegraphics[scale=0.9]{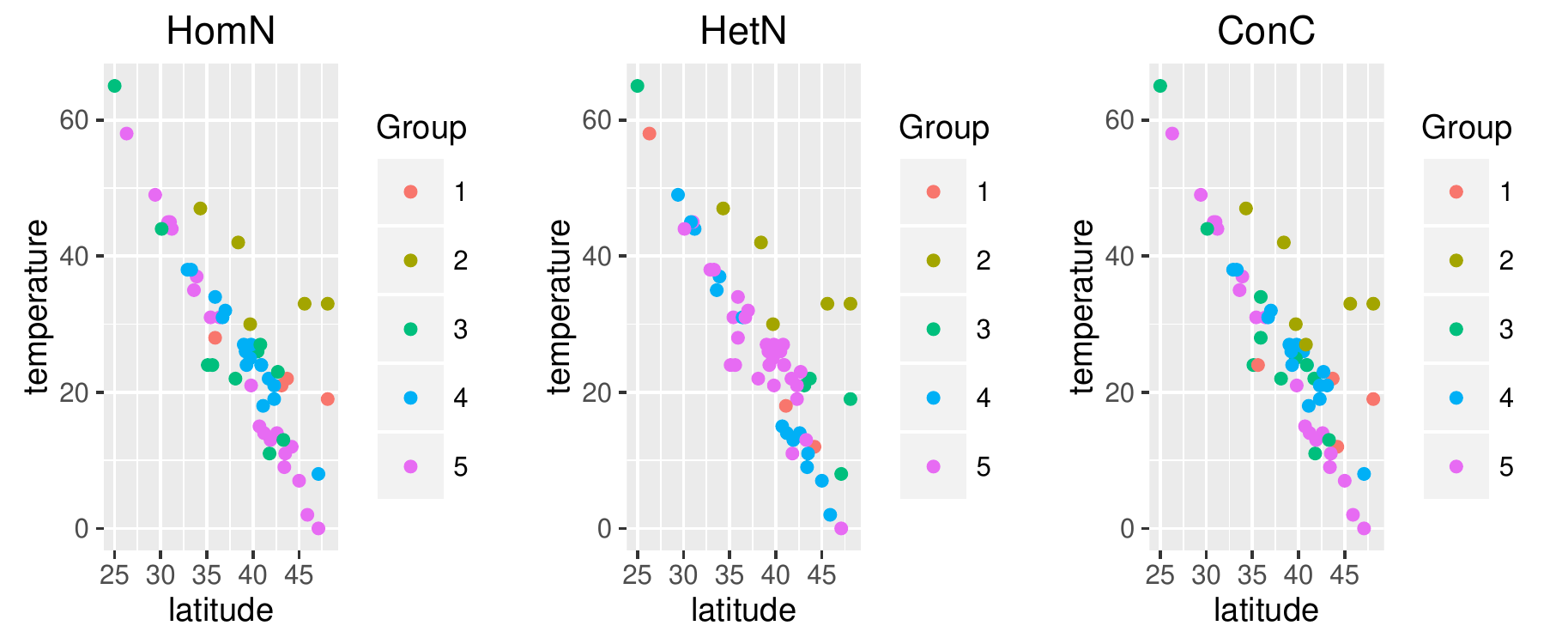}
\caption{\emph{Temperature} data. Best solutions out of 100 random starts, $G= 5$. $K = n/5$, and test set size = $n/10.$}
\label{fig:temp5}
\end{figure}
\end{center}
\FloatBarrier	

The 2-class solution seems to be the most suitable in terms of non-overlapping classes and regression parameters' interpretation. In both classes, latitude has a negative effect on temperature, whereas longitude has a negative effect on temperature in the first (smaller) class, and a positive effect on the second (bigger) class. In the 3-class and the 4-class solutions, the additional classes are mainly obtained from splits of the second (bigger) class: the resulting clusters are characterized by a negative sign for the longitude coefficient. In the 5-class solution, also the first (smaller) class is split into 2 sub-classes, both having the feature of positive sign for the longitude coefficient. In addition, we observe that HetN converged to a spurious solution, which consists in one component having a variance very close to zero. ConC, in all scenarios, estimates a model which is in between HetN and HomN: while being closer to HetN with $G=2$ and $G=4,$ it gets relatively closer to the common scale with $G=3$ and $G=5.$ 

Although we opted for the 2-component solution as the most appropriate one, BIC computed under both HetN and HomN seems to favor the 5-component model (Table \ref{tab:BICtemp}). Due to the spurious nature of the final solution delivered by HetN, $\text{BIC}_{\text{HetN}}$ should not be trusted. We fear this is also the case for $\text{BIC}_{\text{HomN}},$ as the related solution is characterized by a too small class proportion for one component compared to the others ($p_1 = 0.07$), and a small common scale (approximately half of that estimated with $G=4$).

\begin{table}[h!]
\centering
\begin{tabular}{lccccc}
\hline \hline
							&& $G=2$  & $G=3$  & $G=4$  & $G=5$  \\
							\cmidrule{3-6}
$\text{BIC}_{\text{HomN}} $ && 257.98 & 263.33 & 261.81 & 247.06 \\
$\text{BIC}_{\text{HetN}} $ && 257.44 & 256.51 & 251.30 & 107.42 \\
\hline \hline
\end{tabular}
\caption{\emph{Temperature} data. BIC values for $G=2$, $G=3$, $G=4$, and $G=5$, computed under HomN and HetN. Best solutions out of 100 random starts. \label{tab:BICtemp}}
\end{table}

\subsection{Iris data}
We consider first a subset of the Iris data, available at the
link \url{https://archive.ics.uci.edu/ml/index.html}. The data
set contains 3 classes of 50 instances each, where each class refers to a type of iris plant. 
The clusters recovery obtained by the three methods is assessed in terms of Adj-Rand index. We also report computational time and the estimated $c.$  

\FloatBarrier

\begin{table}[h!]
\centering
\begin{tabular}{lcccccc}
\hline \hline
Algorithm && Adj-Rand  && time && $c$\\
\hline
HomN && 0.5532 && 48.5605 && - \\
HetN && 0.4414 && 35.3266 && - \\
ConC && 0.8180 && 495.5422 && 0.0222\\
\hline
\end{tabular}
\caption{\emph{Iris} data. Adjusted Rand index, CPU time and selected $c$ for a 3-component clusterwise linear regression of petal width on sepal width. Best solution out of 500 random starts. $K = n/5$, and test set size = $n/10.$ \label{tab:iris}}
\end{table}

\begin{center}
\begin{figure}[h!]
\includegraphics[scale=0.9]{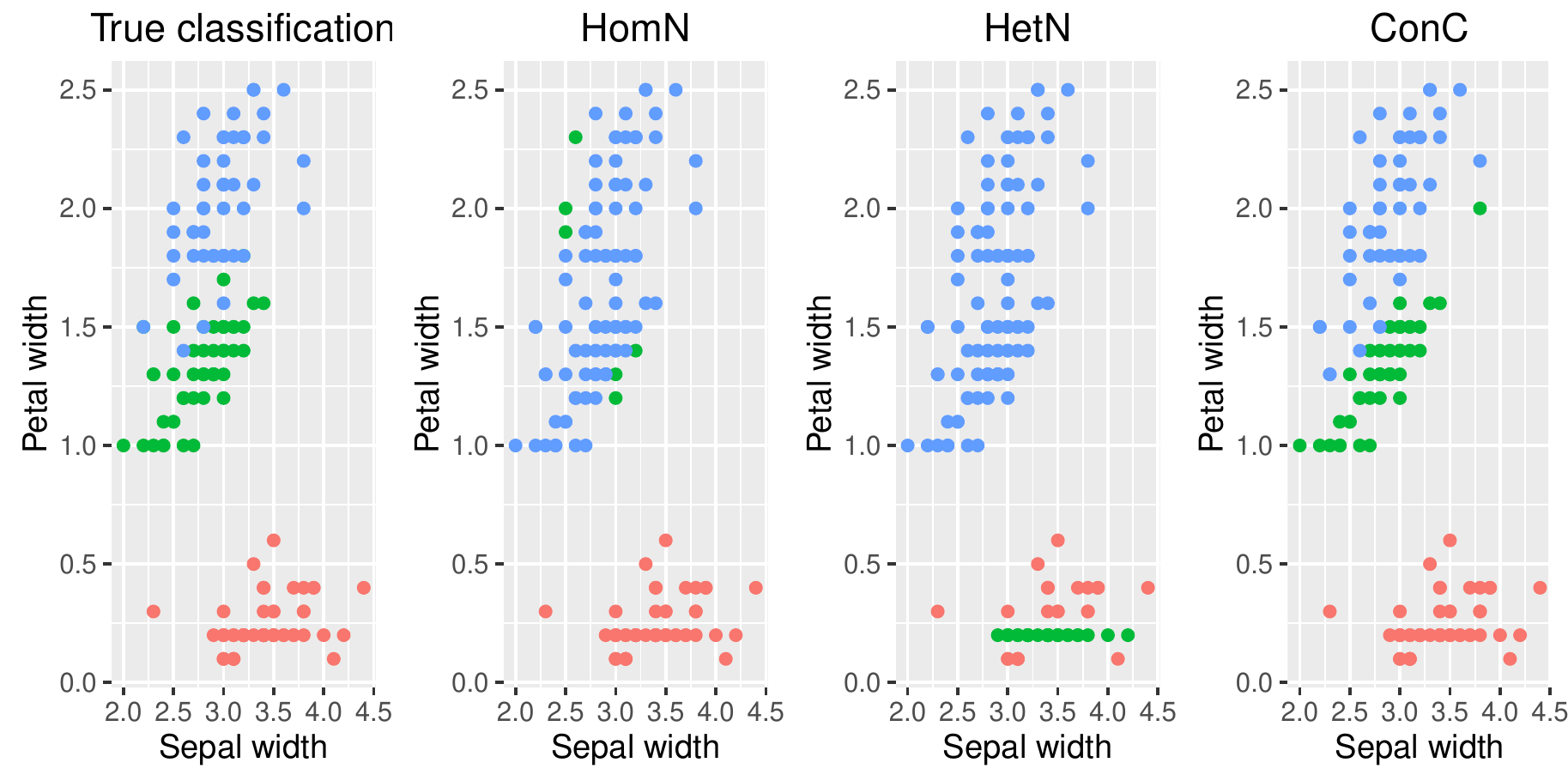}
\caption{\emph{Iris} data. Best solutions out of 500 random starts, $G= 5$. $K = n/5$, and test set size = $n/10.$}
\label{fig:iris}
\end{figure}
\end{center}

\FloatBarrier
ConC yields a clustering which is the closest to the true classification compared to HomN and HetN (Figure \ref{fig:iris}). The Adj-Rand obtained by ConC, as we observe from Table \ref{tab:iris}, is 0.82, whereas HomN and HetN obtain much lower values - 0.55 and 0.44 respectively. On the other hand, the computational time it takes for ConC to run, multiple starting value strategy included, is more than 10 times longer than HetN and HomN.

\section{Conclusions}\label{conclusion} 

In the present paper an equivariant data-driven constrained approach to maximum likelihood estimation of clusterwise linear regression model is formulated. This extends the approach proposed in RGD (2016) for multivariate mixtures of Gaussians to the clusterwise linear regression context. Through the simulation study and the three empirical applications, we are able to show that the method does not only solve the issue of degeneracy, but it is also able to improve upon the unconstrained approaches it was compared with. 

Whereas RGD (2016) showed that the method has merit in both fuzzy and crisp classification, the additional step ahead we take is twofold: 1) we look at how well model parameters are estimated, and 2) we are able to give a strong model interpretation. That is, the value of the selected constant $c$ indicates the most suitable-to-the-data model as a way through homoscedasticity and heteroscedasticity. This does not only relate to the estimated scales, but to the entire set of model parameters and classification, as also the clusterwise linear regressions and clustering estimated with our method correspond to an estimated model in between the homoscedastic model and the heteroscedastic model.

Our method shares common ground with the plain constrained maximum likelihood approach in that parameter updates are the same as in a constrained EM (Ingrassia, 2004; Ingrassia and Rocci, 2007). Nevertheless, the final solution we obtain maximizes the (log) likelihood through a maximization of the cross-validated log-likelihood, and the constraints are tuned on the data. This eliminates all unreasonable boundary solutions standard constrained algorithm might converge to due to the arbitrary way constraints could be put.    

The equivariance property in a clusterwise linear regression framework is related to linear transformations of the dependent variable only. Yet, it is as crucial as in the multivariate mixture of Gaussians case, as it does not uniquely imply that the final clustering remains unaltered as one acts affine transformation on the variable of interest. More broadly, no matter how the data come in, affine equivariance means that there is no data transformation ensuring better results, since the method is unaffected by changes of scale in the response variable.

As it was noticed by Ritter (2014), common scale is highly valuable, but it can be a too restrictive assumption for the clusters' scales. In this respect, our approach does not suffer the inappropriateness of the homoscedastic model, as the constant $c$ controls how close to (or how far from) it the final model will be. Especially in the empirical application, we observed that the method is able to detect departures from homoscedasticity in terms of selected $c.$

The simulation study and the empirical applications have highlighted two open issues. First, the proposed method is computationally intensive compared to the unconstrained approaches it was compared with. Building up a computationally more efficient procedure to select the constant $c$ from the data can be an interesting topic for future research. Second, the BIC computed using the unconstrained models is not always reliable, as we observed from the empirical applications. How to carry out reliable model selection using our equivariant data-driven constrained approach requires further research.


\begin{thebibliography}{123}
\bibitem{arl} Arlot, S., \& Celisse, A. (2010). Cross-validation procedures for model selection. \emph{Statistics Surveys, 4, 40-79}.

\bibitem{}
Bagirov, A. M., Ugon, J., \& Mirzayeva, H. (2013). Nonsmooth nonconvex optimization approach to clusterwise linear regression problems. \emph{European Journal of Operational Research, 229(1), 132-142.}

\bibitem{} Carbonneau, R. A., Caporossi, G., \& Hansen, P. (2011). Globally optimal clusterwise regression by mixed logical-quadratic programming. \emph{European Journal of Operational Research, 212(1), 213-222.}

    
\bibitem{} Chen, J., Tan, X., \& Zhang, R. (2008). Inference for normal mixtures in mean and variance. \emph{Statistica Sinica, 18(2), 443.}

	      
\bibitem{ciuperca} Ciuperca, G., Ridolfi, A., and Idier, J. (2003). Penalized maximum likelihood estimator for normal mixtures. \emph{Scandinavian Journal of Statistics, 30(1), 45-59.}
		
\bibitem{}
Day N.E. (1969). Estimating the components of a mixture of two normal distributions, 
{\em Biometrika, 56, 463-474}.
                        

\bibitem{o} Dempster, A.P., Laird, N.M., and Rubin, D.B. (1977). Maximum Likelihood from Incomplete Data via the EM Algorithm. \emph{Journal of the Royal Statistical Society: Series B (Statistical Methodology), 39, 1-38.}
          
\bibitem{r} DeSarbo, W. S., \& Cron, W. L. (1988). A maximum likelihood methodology for clusterwise linear regression. \emph{Journal of classification, 5(2), 249-282.}


\bibitem{fur} Fr{\"u}hwirth-Schnatter, S. (2006). \emph{Finite mixture and Markov switching models.} Springer Science \& Business Media.


          
\bibitem{}
Hathaway R. J. (1985). A constrained formulation of maximum-likelihood estimation for normal mixture distributions,
{\em The Annals of Statistics, 13, 795-800}.

\bibitem{hng} Hennig, C. (2000). Identifiablity of models for clusterwise linear regression. \emph{Journal of Classification, 17(2), 273-296.}
              
\bibitem{hub1} Huber, P. J. (1967). The behavior of maximum likelihood estimates under nonstandard conditions. In \emph{Proceedings of the fifth Berkeley symposium on mathematical statistics and probability (Vol. 1, No. 1, pp. 221-233).}

\bibitem{hub2} Huber, P.J. (1981). \emph{Robust Statistics.} John Wiley and Sons, New York.

\bibitem{}
Hubert L., \&  Arabie P. (1985). Comparing partitions. {\em Journal of Classification, 2, 193-218}.

\bibitem{} 
Ingrassia S. (2004).  A likelihood-based constrained algorithm for multivariate normal mixture models,
 {\em Statistical Methods \& Applications, 13, 151-166}.

\bibitem{ing07} 
Ingrassia S., \& Rocci, R. (2007). A constrained monotone EM algorithm  for finite mixture of multivariate Gaussians.
{\em Computational Statistics \& Data Analysis,  51, 5339-5351}.


\bibitem{} James, W., \& Stein, C. (1961). Estimation with quadratic loss. In \emph{Proceedings of the fourth Berkeley symposium on mathematical statistics and probability (Vol. 1, No. 1961, pp. 361-379).}

\bibitem{}
Kiefer J., \& Wolfowitz J. (1956). 
Consistency of the maximum likelihood estimator in the presence of infinitely many incidental parameters. {\em Annals of Mathematical Statistics 27, 886–906}.

\bibitem{}
Kiefer, N. M. (1978). Discrete parameter variation: Efficient estimation of a switching regression model.
{\em Econometrica 46, 427-434}.

\bibitem{linsday} Lindsay, B. G. (1995). Mixture Models: Theory, Geometry and Applications. \emph{NSF-CBMS Regional Conference Series in Probability and Statistics, 5, i–163.} 

\bibitem{}
Long, L. H. (ed.) (1972), \emph{The 1972 World Almanac and Book of Facts,} New York: Newspaper Enterprise Association.

\bibitem{}
 McLachlan G.J., \& Peel D. (2000). {\em Finite Mixture Models}. John Wiley \& Sons, New York.
            

\bibitem{}
Peixoto, J. L. (1990). A property of well-formulated polynomial regression models. \emph{The American Statistician, 44(1), 26-30.}
                      
\bibitem{qre} Quandt, R. E. (1972). A new approach to estimating switching regressions. \emph{Journal of the American Statistical Association, 67(338), 306-310.}

               
\bibitem{qre1} Quandt, R. E., \& Ramsey, J.B. (1978). Estimating mixtures of normal distributions and switching regressions. \emph{Journal of the American Statistical Association, 73(364), 730-738}.

\bibitem{} R Core Team (2016). \emph{R: A language and environment for statistical computing.} R Foundation for
  Statistical Computing, Vienna, Austria. \url{https://www.R-project.org/}
	      
\bibitem{} Rocci, R., Gattone, S.A., \& Di Mari, R. (2016). \emph{Under review.}
   
\bibitem{rit} Ritter, G. (2014). \emph{Robust cluster analysis and variable selection.} Monographs on Statistics and Applied Probability 137, CRC Press.

\bibitem{} Smyth, P. (1996). \emph{Clustering using Monte-Carlo cross validation.} In \emph{Proceedings of the Second International Conference on Knowledge Discovery and Data Mining, Menlo Park, CA, AAAI Press, pp. 126–133.}

\bibitem{}
Smyth, P. (2000). Model selection for probabilistic clustering using cross-validated likelihood. \emph{Statistics and Computing, 10(1), 63-72.}

\bibitem{st} Stone, C. (1984). Cross-validatory choice and assessment of statistical predictions. \emph{Journal of Royal Statistical Society: Series B (Statistical Methodology), 36(2), 111-147}.
	      

\end{thebibliography}
\end{document}